\documentclass[a4paper, 10pt, conference]{ieeeconf}      % Use this line for a4
\IEEEoverridecommandlockouts                              % This command is only
\usepackage{epsfig} % for postscript graphics files
\usepackage{amsmath} % assumes amsmath package installed
\usepackage[nocomma]{optidef} 
\usepackage{verbatim}
\usepackage{amssymb}  % assumes amsmath package installed
\usepackage{graphicx}
\usepackage{float}
\usepackage{subcaption}
\usepackage{caption}\usepackage{soul}
\usepackage[linesnumbered,ruled,vlined]{algorithm2e}
\usepackage{listings,multicol}
\usepackage{lipsum}
\usepackage{setspace}
\usepackage{multirow}
\usepackage[table,xcdraw]{xcolor}
\usepackage{color}
\usepackage{textcomp}
\usepackage{adjustbox}
\usepackage{nomencl}
\usepackage{multicol}
\usepackage{booktabs, multirow} % for borders and merged ranges
\usepackage{soul}% for underlines
\usepackage[table]{xcolor} % for cell colors
\usepackage{changepage,threeparttable} % for wide tables
\usepackage[compact]{titlesec}         % you need this package
\titlespacing{\section}{0pt}{0pt}{0pt} % this reduces space between (sub)sections to 0pt, for example

\usepackage{textcomp} 
\definecolor{listinggray}{gray}{0.9}
\definecolor{lbcolor}{rgb}{0.9,0.9,0.9}

\newcommand*{\affaddr}[1]{#1} % No op here. Customize it for different styles.
\newcommand*{\affmark}[1][*]{\textsuperscript{#1}}
\newcommand*{\email}[1]{\texttt{#1}}
\usepackage[a4paper, total={184mm,239mm}]{geometry}
\def\BibTeX{{\rm B\kern-.05em{\sc i\kern-.025em b}\kern-.08em
    T\kern-.1667em\lower.7ex\hbox{E}\kern-.125emX}}

\usepackage{amsmath}
\begin{document}
\title{\textbf{Co-designing Intelligent Control of Building HVACs and Microgrids}}

\author{%
Rumia Masburah$^{*,1}$,
Sayan Sinha$^{*,1}$, Rajib Lochan Jana$^{*,1}$,  Soumyajit Dey\affmark[1], and Qi Zhu\affmark[2]\\
\affaddr{\affmark[1]\textit{Indian Institute of Technology Kharagpur}}, 
\affaddr{\affmark[2]\textit{Northwestern University}}\\
{\{rumiamasburah, sayan.sinha, rajib.jana\}@iitkgp.ac.in},
soumya@cse.iitkgp.ac.in\\
qzhu@northwestern.edu
\vspace*{-0.1in}
\thanks{{$^{*}$ denotes equal contribution. }Ack: The work was funded by MHRD and Department of Power, Govt. of India, under the IMPRINT project no. 6158.}
}
\maketitle

\begin{abstract}
Building loads consume roughly 40\% of the energy produced in developed countries, a significant part of which is invested towards building temperature-control infrastructure. Therein, renewable resource-based microgrids offer a greener and cheaper alternative. This communication explores the possible co-design of microgrid power dispatch and building HVAC (heating, ventilation and air conditioning system) actuations with the objective of effective temperature control under minimised operating cost. For this, we attempt control designs with various levels of abstractions based on information available about microgrid and HVAC system models using the Deep Reinforcement Learning (DRL) technique. We provide control architectures that consider model information ranging from completely determined system models to systems with fully unknown parameter settings and illustrate the advantages of DRL for the design prescriptions.

%Building microgrids provide a useful technique for renewable based energy dispatch which helps in.... Unlike the existing works, in this paper, we present Deep Reinforcement learning (DRL) techniques for microgrid control for optimising the electrical expenses of building networks by co-optimising the control actions for microgrid and building heating, ventilation and air conditioning (HVAC) system, while maintaining the comfort need of the building occupants. Here, we illustrate the DRL techniques for partially known and completely unknown system models and also compare them with a complete model-based system model. In this work, the AI based techniques are used for analysing the energy consumption data, predicting future energy demand and establishing a suitable energy saving policy.

\end{abstract}

%%%%%%%%%%%%%%%%%%%%%%%%%%%%%%%%%%%%%%%%%%%%%%%%%%%%%%%%%%%%%%%%%%%%%%%%%%%%%%%%

%\textbf{\textit{Index Terms----}Microgrid, Building Energy Management System, HVAC controller, Deep Reinforcement learning, DDPG Method, MDP.}

%\makenomenclature
%\mbox{}
%\nomenclature{$e_t^{net}$}{Net Power Demand}
%\nomenclature{$e_t^{const}$}{Constant Load}
%\printnomenclature

% \maketitle
\section{Introduction}
Microgrids with renewable power sources have emerged as a viable alternative to the growing energy demand (about 40\%) \cite{energy} of building loads in developed countries. In large smart building loads, the heating, ventilation, and air conditioning (HVAC) systems account for the highest percentage ($70\%$)  of the overall building load \cite{bd1}. Given this huge energy requirement, an important aim of Building Energy Management Systems (BEMS) is to optimise the HVAC energy consumption, while maintaining the comfort needs of building occupants. This has an immediate impact on energy cost expense reduction for the building. 

\par Recent research in the area of electrical cost minimisation for a network of smart buildings can be broadly divided into two major classes. One direction of research focuses on optimal control of building microgrids to decrease dependency on the utility grid and balance energy supply and demand by utilising storage systems efficiently \cite{d2,d3}. The other direction focuses on the optimal control for building HVAC systems and load management by the BEMS to optimise the building energy consumption for reducing electricity costs \cite{afram2014theory, fan_power}. Thus one may observe that while intelligent control of microgrid systems reduces the energy dependency for building networks (i.e. a collection of community building loads) on the utility grid, the BMS tries to optimise {\em in-building} actuators for reducing building power-demands. These two avenues of power dispatch, control and system operations can be made to operate more efficiently if cross-layer, coordinated optimisations are enabled among them. This idea forms the premise of our work. The operational cost optimisation of a building network depends on the supply from the microgrid, demand of BEMS and real-time price of the utility grid, and different operational constraints imposed by the microgrid and BEMS. So the control actions for the two systems, i.e. microgrid and BEMS, may be intertwined for reducing the total electrical cost. It leads to co-designing the control of the two systems (microgrid and the BEMS). %Such optimisation problem involves multi-variables with constraints.

Model predictive control (MPC) shows its efficiency for optimal control design with underlying complex constrained multivariable optimisation problems in many domains \cite{afram2014theory}. The MPC-based strategies require accurate prediction on the power consumption of building loads like HVACs, which depend on the accuracy of building thermal models \cite{d5}. Developing an accurate building thermal model from scratch is a time-consuming, and costly process \cite{afram2014theory}. It is often intractable to get accurate dynamics of the temperature inside the building, which is affected by many factors. Moreover, thermal models are building-specific due to heterogeneity present in building design, the material used and operations. For this reason, an alternative control method is to use learning-based algorithms that directly learn the control strategies from sensor data. However, learning-based techniques require a significantly large,  effective and diverse dataset depending on building size and internal structure. In the case of building models, such datasets are not readily available \cite{bd2}.  %The dataset is different for every other building due to structural or operational difference.

In order to address these issues, Reinforcement Learning (RL) \cite{RL} has been used as a possible real-time control technique in building scenarios. Leveraging   Deep Reinforcement learning (DRL) for continuous large datasets \cite{a11} generates near-optimal actions without having a supervisor, via trial-and-error paradigm, for handling a complex control problem. DRL based HVAC control works \cite{transfer} mainly focus on calculating energy schedules in BEMS using predicted day-ahead data rather than real-time experience and information.  In retrospect, the present work is the first to apply DRL technique for co-optimising the control actions for microgrid and HVAC systems with the objective of minimising the electrical expenses of building networks while maintaining the comfort needs of the building occupants at different thermal zones. 

However, such co-optimisation of two structurally independent but functionally related systems involves a very large state-space, making parameter learning and real-time control challenging. Based on this observation, the present work analyses three different possible state information and control scenarios and performs the co-optimisation.
\textbf{{\em Scenario-1}}: Parameters of grid components and the building thermal models are all available. In this case, we employ an integrated MPC based coordinated control of the grid and building loads. \\
\textbf{{\em Scenario-2}}: Parameters of the grid components are not known, but the building thermal model is available with an MPC based temperature control. In this case, we design a DRL based microgrid controller, which additionally supervises the parameters of the HVAC MPC. \\
\textbf{{\em Scenario-3}}: All the parameters of the grid, as well as the building thermal models, are unavailable. So, we design a pure DRL based approach for both grid and HVAC control surfaces. We formulate the underlying sequential decision making problem as an MDP (Markov Decision process) with suitable environment states, actions and reward function.
%often leads to a situation when a partial model of the overall system, i.e. mathematical models for different parts of the systems, are available. In that case, we show how to apply the DRL method for control design of the overall system incorporating the known mathematical model of one system. We also, show the use of DRL when the model is entirely unknown.
The main contributions of our work are summarised as follows:
\begin{itemize}
    \item  We propose a framework for co-designing the control operations of a microgrid system with building loads. The control operations include the energy storage system charging and discharging, power exchange between the microgrid and the utility grid, real-time price of utility grid, comfortable temperature range for the building, HVAC power adjustment and weather uncertainties. 

%\item We design a control strategy for joint operation of the microgrid energy management system (MEMS) and the HVAC controller based on Deep Deterministic Policy Gradient (DDPG) policy. 

\item We solve the control problem assuming three different abstraction levels as discussed earlier. In that way, we provide a systematic approach for the application of DRL techniques in this problem, assuming different levels of system parameter information.  We compare the performance of the proposed DRL-based algorithms ( \textit{Scenario-2 and 3}) with the  pure model-based control approach (\textit{Scenario-1}). %We also discuss the pros and cons for each of these different levels of abstraction.
\end{itemize}
\section{System Description and Problem Statement}
\begin{figure}
\centering
\resizebox{\linewidth}{!}{
\includegraphics[scale=0.10]{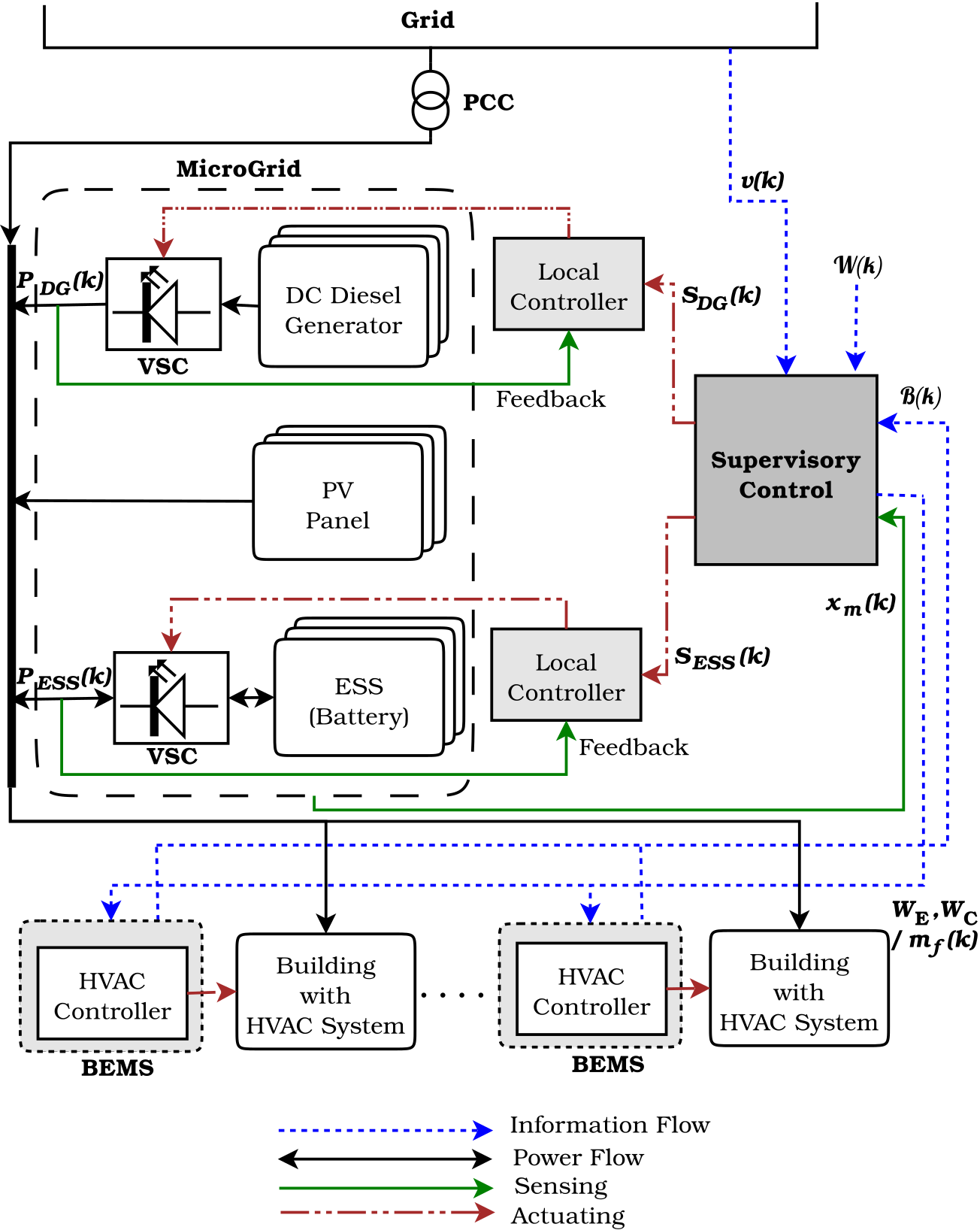} 
}
\caption{Control Framework} 
%\vspace*{-0.4cm}
\label{fig:framework}
\end{figure}

The system model for this work is depicted in Fig. \ref{fig:framework}. It comprises a set of building electrical loads and a microgrid (coupled with the utility grid) that supplies the electricity to the building loads.  Brief description of the different components of our studied system are illustrated in this section.
\subsection{Microgrid System}
The microgrid consists of PV panels as uncontrollable generation units, DC diesel generators as controllable generation units and batteries as the electrical storage system (ESS). The units are connected to the microgrid through controllable voltage source converters (VSC). We adopt the standard method for VSC control with inner loop current controller and outer loop voltage controller, which regulates the VSC output voltage to a desired reference \cite{VSC}.

% \begin{figure}
% \centering
% \resizebox{\linewidth}{!}{
% \includegraphics[scale=0.05]{fig/system_fram.eps} 
% }
% \caption{Schematic diagram of Studied System} 
% %\vspace*{-0.4cm}
% \label{fig:system}
% \end{figure}
Let us assume, the sampling interval of the microgrid is $t_s$. At some instance $k$, the PV power generation is governed by the following dynamical equation \cite{c7}, 
\begin{equation}\label{g1}
    P_{solar}(k)=A*y*r(k)*R
\end{equation} 
where, we denote $A$ as the total PV panel area,  $y$ as the solar panel yield, $r$ as the solar  radiation and $R$ as the performance ratio. For the storage system i.e. the battery model, the charging/discharging dynamics of the battery is given as,
\begin{equation}\label{g2}
\begin{split}
SOC(k+1) = (1-\delta_{ESS}t_s)SOC(k)-u_{ESS}(k)\frac{P_{ESS}(k)}{ \eta_{dis}{P^{max}_{ESS}}}\\ 
+ (1-u_{ESS}(k))\frac{P_{ESS}(k)}{ \eta_{ch}{P^{max}_{ESS}}}  
\end{split}
\end{equation}
 following  \cite{c7}, where $SOC(k), P_{ESS}(k)$ are the  current state of charge (SOC) of the battery and the power supply from/to the battery during discharge/charge respectively. Also $\delta_{ESS}$ is the  self-discharge rate of storage. The symbol $\eta_{dis}$/$\eta_{ch}$ represents discharging/charging efficiency and  $P^{max}_{ESS}$ is maximum storage capacity of the battery. The $u_{ESS}$ is the charging/discharging command, $u_{ESS}(k)=1$ indicates the discharging mode and $u_{ESS}(k)=0$ indicates the charging mode. %The charging and discharging modes are mutually exclusive for the battery system.
Power output dynamics ({$P_{DG}$}) of the diesel generator can be given as \cite{c6},
\begin{equation}\label{g4}
{P}_{DG}(k+1) =-\frac{1}{\tau_{DG}}P_{DG}(k)+\frac{u_{DG}(k)}{\tau_{DG}}
  \end{equation}
where, $\tau_{DG}$ is the average delay between the delivery and power command  and $u_{DG}$ is the power commanded to the  generator by the controller. 

% The discretized overall power generation dynamics of microgrid can be written as given in the following equation, 
% \begin{align}\label{eq:micro}
%     &x_m(k+1)=A_mx_m(k) + B_mu_m(k) + E_md_m(k)\\
%     &y_m(k+1)=C_mx_m(k+1)\nonumber
% \end{align}
% where, $A_m$, $B_m$,$C_m$ and $E_m$ are state, input, output and disturbance matrices respectively, defined using Eq. (\ref{g1} - \ref{g4}). The state $x_m(k)$, input $u_m(k)$, output $y_m(k)$ and disturbance $d_m(k)$ vectors can be defined as follows,
% \begin{equation}
% \begin{split}
% x_m(k) \stackrel{def}{=} &[SOC(k)\, P_{solar}(k)]^T\\
% u_m(k) \stackrel{def}{=} &[u_{ESS}(k)\, P_{ESS}(k)\, P_{DG}(k)]^T\\
% y_m(k) \stackrel{def}{=} &[SOC(k)\, P_{solar}(k)]^T\\
% d_m(k) \stackrel{def}{=} &[D_{SOC}(k)\,  D_{P_{DG}}(k)\,  h(k)]^T \\
% \end{split}
% \end{equation}
% The terms $D_{SOC}(k),\,  D_{P_{DG}}(k)$ are the model disturbances in respective state variables and $h(k)$ is the forcasted solar radiation. 
\subsection{HVAC system \& Control Framework}
In this work, we consider each building of the building network to be equipped with an HVAC system. We use the hierarchical economic Model Predictive Control (eMPC) framework from \cite{d5} for HVAC systems to satisfy the comfort need of the in-building environment while minimising the energy consumption. The higher-level eMPC of HVAC calculates the optimal airflow rate into each zone. The input to the building for temperature control is, in fact, the air mass flow rate that should enter each currently used thermal zones through the ducts. These optimal input values are given to the lower-level controllers (PID controller) as the setpoints for air mass flow. The output of the lower-level controllers is basically control signals that adjust the angle of dampers which regulate the amount of air blown into the thermal zones. In this work, we are interested in the electrical load of the building that affects the control of microgrid power dispatch.
%\textcolor{red}{one review: The modelling of the HVAC system is  cursory. Such systems are complex and consist of air handling units, variable airflow fans, and recirculation units. Each of them has a separate timescale of control. For example, the air handling unit temperature setpoint can only be changed approximately once in an hour. This complexity is not reflected in the modelling used in this paper. In fact, only one fan is modelled in the paper, which is far from reality.}
 We assume that other electrical loads  except HVAC are constant. %The power requirement of HVAC as controllable load is made part of our formulation, though the methodology can also accommodate other loads in similar manner. 
 The power requirement of the HVAC system is mainly governed by the power requirement of the variable speed driven fan and chiller. The power demand of the fan can be modelled as follows \cite{d5}:
\begin{equation} \label{eq:7}
 P_f(t) = k_f (\dot{m_i})^2
\end{equation}
where $k_f$ is a parameter that captures both the fan efficiency and the duct pressure losses while $\dot{m_i}$ is the supply air mass flow rate of the fan.
The chiller power demand can be modelled as
\begin{equation} \label{eq:8}
 P_c(t) = \dfrac{c_p}{\eta} \dot{m_i}(T_s -T_c)
\end{equation}
following  \cite{d5} where, $\eta$ is coefficient of performance for chiller, $c_p$ is specific heat capacity of air, $\dot{m_i}$ is the supply air mass flow rate, $T_s$ is the temperature of air entering the chiller and  $T_c$ is the temperature of air going out from the chiller.

Let us assume, prediction horizon and sampling period are $H^p$ and $t_s$, respectively. In order to calculate the optimal airflow rate, the higher-level eMPC in \cite{d5} solves an optimisation problem at each sampling time $k$. The objective of the problem is to trade off the energy consumption and comfort of the building that is captured in the following objective function.  
\begin{equation} 
 w_E E_H(k) + w_c \sum_{i=0}^n CR_i * \dfrac{1}{CF_i(k)}
\end{equation}
where, $E_H(k)=\int_{kt_s}^{(k+H_p)t_s} [P_f(t) + P_c(t)]dt$ is the energy consumption of the HVAC used in \cite{d5} within the time interval $[kt_s, (k+H_p)t_s]$. The term $CF_i(k) = \sum_{j=k}^{k+H_p} \dfrac{1}{|T^{desire}_i - T_i(j)|}$ is the comfort factor of $i^{th}$ zone ($1\leq i \leq n$) within the time interval $[kt_s, (k+H_p)t_s]$, where $T^{desire}_i$ is the desired temperature (set point) and $T_i(j)$ is the  temperature of the $i^{th}$ zone at $j$-th sampling instant.The comfort factor implies the deviation of current temperature from the set point and the higher value of it indicates that the zone temperature is closer to the desired value. The term $CR_i \in (0,1)$ indicates the air conditioning criticality of the $i^{th}$ zone. The criticality of zones sets the relative priority for maintaining high comfort among the zones. The $w_E, w_c \in (0,1)$ are the weight parameters for energy consumption and comfort need respectively. The   energy consumption of the HVAC system depends on the choice of system parameters $w_E$ and $w_c$. These two parameters are needed to be suitably tuned for deciding the trade-off between the relative priorities of comfort and energy consumption.

\section{Proposed HVAC-aware MEMS Framework}
In this section, we describe our proposed control framework, depicted in Fig. \ref{fig:framework} for the microgrid energy management system (MEMS). We use a two-level hierarchical control design, similar to \cite{Morstyn}, for optimally managing the microgrid operation to minimise the electrical expenses of a building network. At each sampling time $k$, our high-level supervisory controller senses the state $x_m(k)$ of the microgrid and generates the optimal schedule for the ESS and diesel generators (DGs) of microgrid along with a suitable choice of HVAC system input $\mathcal{H}(k)$. The schedule for the storage, $\mathcal{S}_{ESS}(k)$, indicates the amount of power to be fed to or drawn from the microgrid for a specific time duration $[k, k+1]$ during discharging or charging respectively. Similarly, the schedule of diesel generators $\mathcal{S}_{DG}(k)$ indicates the amount of power to be fed to the microgrid for the duration $[k, k+1]$. The input $\mathcal{H}(k)$ of HVAC system is either parameters $w_E$, $w_c$ or control inputs $\dot{m}$ of HVAC. As mentioned earlier, the $w_E$ and $w_c$ are the design parameters for HVAC control, and the energy consumption of the HVAC system depends on the choice of them \cite{d5}. As a result, the suitable values for $w_E$ and $w_c$ are chosen by the supervisory controller to minimise the electrical expenses and the model-based HVAC control system generates the control inputs based on these choices of values. Therefore, the supervisory controller generates $w_E$ and $w_c$, when the model of HVAC is known and generates control inputs $\dot{m}$ of HVAC when the model of HVAC is unknown. The operation of the supervisory controller can be formally described by
\begin{equation}
    \{\mathcal{S}_{ESS}(k), \mathcal{S}_{DG}(k), \mathcal{H}(k)\} = 
    f(x_m(k), \mathcal{B}(k), \mathcal{W}(k), v(k))
\end{equation}
%\sd{no eq no. also what is /mf in LHS ?? not clear to review}
where, $\mathcal{B}(k), \mathcal{W}(k), v(k)$ represent the power demand of the building network, weather data forecast and real-time price of grid respectively for the interval $[k, k+1]$. 
% The generated schedule for ESS and DGs are fed to the respective lower-level local controller as the setpoint of voltage references for the interval $[k, k+1]$. The output of the lower-level controllers is basically controlling signals for VSCs to regulate the power flows in the microgrid. Therefore, the ESS and DGs are controlled locally by lower-level controllers which track the setpoint given by the higher-level controller.  

For designing this supervisory controller, we explore a DDPG-based DRL technique and reduce the dependability of the historical data. %We define the MPC based HVAC model to be known as \textit{partially known system} and models for both the subsystem to be unknown for \textit{completely unknown system}. \sd{senetcne has no meaning} 
We describe a systematic approach for applying such a learning technique to design the MEMS supervisory controller for Scenario-2 and 3, for achieving the optimisation goal, i.e. the co-optimisation of two subsystems (microgrid and the building HVAC system). %\sd{why}

\section{DRL Based Controller Design for Partially Known System model}
 In microgrids, 
% as it is very hard to track the statistical distributions of all the system parameters like renewable generation output, power demand and electricity price. Hence 
the operations of ESS and HVAC systems are intertwined, which means that the current decision in any of the systems would affect the future decision of the other system. So, we formulate the sequential decision making problem in the microgrid as an MDP (Markov  Decision Process). %\sd{why} 
% In building loads, the comfort factor at next time slot is dependent on the indoor temperature, HVAC power input, and environment disturbances (e.g., outdoor temperature and solar irradiance intensity) in the present time slot. In this study, DRL based method is applied to microgrid for calculating the optimal cost effective operation schedule of the generating units like diesel generator, energy storage device and utility grid. Here, we also try to optimise the weight parameters of MPC-based HVAC controller in order to decrease the electrical demand further resulting in reduction in electricity bill by keeping the comfort factor within desirable range. 
In this case, instead of using the air mass flow rate for regulating the temperature, we predict the values of the MPC parameters($w_E$ and $w_c$) within a range for minimising the electrical consumption of the building load further and also maximising the comfort factor.

\textbf{MDP Formulation:} A MDP is formally defined as a four-tuple $M = (s, a, P, R)$, where $s$ is the set of environment states, $a$  is the set of actions, $P$ is the state transition probability function and $R$ is a reward function.

In this work, the agent is defined as the learner and decision-maker (i.e., microgrid with HVAC controller), while the environment is comprised of objects like renewable generators, ESS, HVAC system, utility grid, indoor/outdoor temperature etc. The agent observes environment state $s_t$ and takes action $a_t$. Then, environment state becomes $s_{t+1}$ and the reward $R_{t+1}$ is returned. In the following parts, we will design the key components of the MDP, including environment state, action and reward function.\\
The optimal solution of the given problem can be solved by observing the current system states. If $SOC$ denotes the State of Charge (SOC) of ESS and $P^{max}_{ESS}$ is the maximum capacity of ESS. For preventing overcharging and over-discharging of ESS, we introduce a factor $\gamma$ as a safety factor. Then we can write 
\begin{equation}
\gamma.P^{max}_{ESS} \leq SOC \leq (1-\gamma).P^{max}_{ESS} 
\end{equation}
We define the net power demand $P_{net}^t$ of the building load (assuming HVAC load $P_{H}^t$ as the controllable load and other loads $P_{const}^t$ as constant) at a time step $t$ as
\begin{equation}
P_{net}^t = P_{H}^t + P_{const}^t - P_{solar}^t
\end{equation}
We also assume that there is no loss in ESS while charging, discharging or staying idle. $P_{solar}^t$ is the predicted PV power for time step $t$.
The time is repeated with a period of 24 hours, the state of the time stamp($\widehat{t}$) is considered to be repeated over 24 hours.

\textbf{Environmental State:} We denote the real-time or dynamic electricity price of the utility grid as $v_t$ and the parameters related to two contradictory factors (energy consumption and comfort) of the MEMS as $w_E$ and $w_c$.
The state of the Microgrid with the building load can be written as \quad $s_t= [SOC^t \quad P_{net}^t\quad  v_t\quad  T_{in} \quad T_{out} \quad w_c \quad \widehat{t}] \in $ $S$  where $S$ is the state space of the energy management system of the building.
%\vspace{0.6 mm}

\textbf{Action Space:} Let us denote the power exchange between the microgrid and utility grid as $P_{ut}$ and ESS charging/discharging power as $P_{ESS}^t$. The range of $P_{ESS}^t$ is [$-d_{max},c_{max}$] where $d_{max}$ is the maximum discharging power and $c_{max}$ is the maximum charging power. 
For maintaining the power balance, the power supply should be equal to the power demand.
Then we can have 
\begin{equation}
P_{ut} + P_{solar}^t - P_{ESS}^t = P_{const}^t + P_{H}^t
\end{equation}
Again we know $P_{H}^t$ is a function of $w_E$ and $w_c$. For reducing the overall electrical consumption, we have to tune these two parameters. Let the action related to the tuning parameters be $T_p$. In this case, the action space $A$ is:
$A=\{P_{ESS} \quad T_p \}$.

\textbf{Reward:} Let $R(s_t,a_t)$ be the reward function that gives the cost of the energy that the microgrid sell to the utility company. The reward function can be defined as :
\begin{align}
 R(s_t,a_t) & = r^b+r^{net} \quad\text{where}, \\
   r^b&=-[SOC^t.u_t+\epsilon.(P^{max}_{ESS})],\quad SOC^t\leq \gamma.P^{max}_{ESS} \label{eq:rba}\\
  r^b&=-[SOC^t.u_t+(1-\epsilon).P^{max}_{ESS}], SOC^t\geq \gamma'.P^{max}_{ESS} \label{eq:rbb}\\
  r^{net}&=-[P_{net}^t.u_t+\kappa^{up}(w_c-w_c^{max})],\quad w_c>w_c^{max}\label{eq:c3}\\
  &=-[P_{net}^t. u_t+\kappa^{low}(w_c^{min}-w_c)],\quad w_c<w_c^{min}\\
  &=-[P_{net}^t. u_t],\quad otherwise.
  \end{align}
  Here $\epsilon$ is the penalty factor for overcharging or overdischarging the ESS and $\kappa^{up}$, $\kappa^{low}$ are penalty factors for the MPC based HVAC parameter values if they are out of range and $\gamma'=(1-\gamma)$. The values of the parameters are in the range [0,1]. Selling electricity price $u_t$ can be defined as $u_t=\sigma v_t$ where $\sigma$ is a constant.\\
 \textbf{Action-Value Function:}
\vspace{0.6 mm}
 The estimation of the function $f$ illustrates the relationship between the parameters, the total power consumption $P_{total}$ and comfort factor:
\begin{equation}
\widehat{f}(w_c,w_E,P_{t-1}^{total})=P_{t}^{total}
\end{equation}
where $\widehat{f}$ represents the approximated function that explains the relationship between the input data.  

\section{DRL Based Controller Design for unknown System model}
In this section, we try to optimise the MEMS with renewable power sources, energy storage system, HVAC systems without having any knowledge about the system. To be explicit, our goal is to minimise the electricity cost by maintaining a comfortable indoor temperature. However, due to the unavailability of accurate dynamics of indoor temperature, which is affected by many parameters \cite{bd1}, it is tough to track the statistical distributions of all the random system parameters like renewable generation output, outdoor temperature and electricity price. 
% Here we have coupled constraints associated with the operations of ESS and HVAC systems which means that the current action would affect the future decisions.To address the above concerns, we propose a DDPG based algorithm which makes decisions about ESS charging/releasing power and HVAC control input, essentially dependent on the current information.

Unlike the partially known system method, here we consider the whole building network along with the microgrid to be unknown along with HVAC controller dynamics. This motivates the case of a pure DRL based approach where a DRL agent is responsible for controlling both the HVAC control surfaces and the microgrid settings. 
% The agent explores the state space and incrementally learns different operating scenarios in a given environment.
% \vspace{0.8mm}

\textbf{MDP Formulation:}
% A prerequisite for utilising RL in building control is to structure the control problem as a Markov decision process (MDP) as mentioned earlier.
Many control tasks like building energy management systems, having the stochastic idea of the environment (e.g., occupants, weather) frequently include continuous actions (e.g., temperature setpoint, supply air mass flow rate). A mere discretisation of the continuous action space rapidly gets intractable because of the dimensionality problem \cite{dim}. As discussed earlier, we present a Deep Deterministic Policy Gradient (DDPG) based DRL model to represent the controller of the HVAC system.
%As stated earlier, the primary motive of the controller is to determine the air mass flow rate of the system. We control the air mass flow rate of every room independently. In this direction, we execute similar but separate models for every room. This also assists in enhancing the scalability of the model. 
The Markov Decision Process (MDP) model for the DDPG is given as explained below.:
\vspace{0.8mm}
\begin{figure}[h]
\centering
\resizebox{\linewidth}{!}{
\includegraphics[scale=0.10,clip]{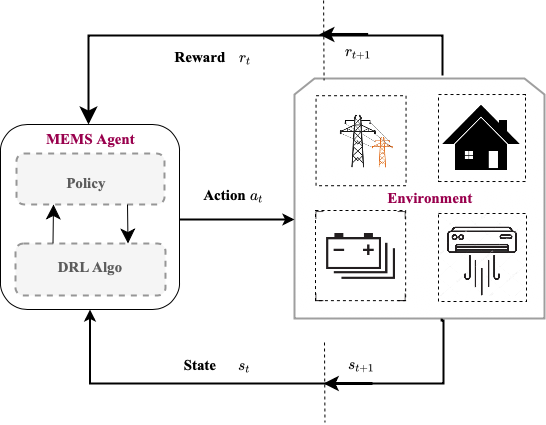} 
}
\caption{The agent-environment interaction in the MDP.} 
%\vspace*{-0.4cm}
\label{pure}
\end{figure}
\vspace{0.8mm}

\textbf{Environmental State:}
The state space $S$ of the MDP can be represented as a five dimensional vector $s_t= [SOC^t \quad  v_t \quad T_{all}(i)  \quad T_{\delta} \quad T_{out}]$ at a given time step $t$. Here, $e_t^b$ is the amount of charge on the battery, $T_{all}(i)$ is the temperature of all the zones for ith room and $T_{\delta}$ is the difference in temperature from the comfortable range and %Do note that this state space is for every room. 
$T_{out}$ is the outside air temperature.
The vector $T_{all}$ is cycled in an orderly fashion depending on the room it is representing such that the temperature of the current room is represented by the first dimension. For instance, if for room $1$, $T_{all}(1)$ is given by $[T_1 \quad T_2 \quad T_3 \quad ... \quad T_n$], then for room $2$, the same would be given by $[T_n \quad T_1 \quad T_2 \quad ... T_{n - 1}]$. Here $n$ is the total number of rooms.

\textbf{Action Space:}
The action space $A$ for the model is given by  $a_t$= $[ P_{ESS}^t,\quad $\.{m}$]$, where $P_{ESS}^t$ at a given time step $t$, is the amount of charging/discharging power of the battery and \.{m} is the output air mass flow rate.

% We propose two possible reward functions here, the first one being given by 
% \begin{equation}
% r^b + r^{room}
% \end{equation}
% Here,
% \begin{equation}
% r^{room} = r^{net} \dfrac{\dot{m}_i}{\sum_i \dot{m}_i}
% \end{equation}
% The formulation for $r^{room}$ is derived from \cite{bd1}. $r^{b}$ and $r^{net}$ are the same as given in Equation-$14$ and $15$ respectively. Here $m_i$ is the air mass flow rate of the $i^{th}$ room.
\textbf{Reward:}
The reward function is defined as 
\begin{equation}
r = - (power\textunderscore cost) - \lambda T_{\delta}
\end{equation}
where $power\textunderscore cost$ is the cost of the power given the current pricing scheme. The cost is taken to be directly proportional to the air mass flow rate. $T_{\delta}$ is the difference in temperature of the room and the comfortable range of temperature and $\lambda$ here is a hyper-parameter. 
% We present the results using reward function presented in Equation-$20$, which gives better results compared to the other one.

\textbf{Action-Value Function:}
While co-optimizing the ESS and the HVAC system at time step t, the MEMS agent intends to maximize the total reward collected over the future \cite{RL}. Let $Q(s,a)$ be the action-value function under the DDPG policy $\pi$, which represents the expected return if action $a_t = a$ is taken in state $s_t = s$ under the policy $\pi$ . Then, the optimal action-value function $Q^{\ast}(s,a)$ is max $Q(s_t,a_t)$ which can be evaluated using Bellman equation, i.e.,
\begin{align}
 Q^{\ast}(s, a)&=\mathbb E[r_{t+1} + \gamma max_{a\textquotesingle} Q^\ast(s_{t+1}, a\textquotesingle)|s_t = s, a_t = a]\\
 &=\sideset{}{_{s\textquotesingle,r}}\sum P(s\textquotesingle,r|s,a)[r+\gamma max_{a\textquotesingle} Q^\ast(s\textquotesingle,a\textquotesingle)
 \end{align} 
 where $s\textquotesingle \in \mathcal{S}, r \in \mathcal{R}, a\textquotesingle \in \mathcal{A} $ and $P \in \mathcal{P}$.\\
The state transition probabilities $P(s\textquotesingle,r|s,a)$ are required for evaluating $Q^\ast(s,a)$. Due to available noises and disturbances, it is difficult to obtain these transition probabilities accurately as mentioned earlier. To support the case with continuous system states, we use DDPG based algorithm.

\begin{figure*}
\centering
% \resizebox{\linewidth}{!}{
\includegraphics[width=\textwidth,height=9cm]{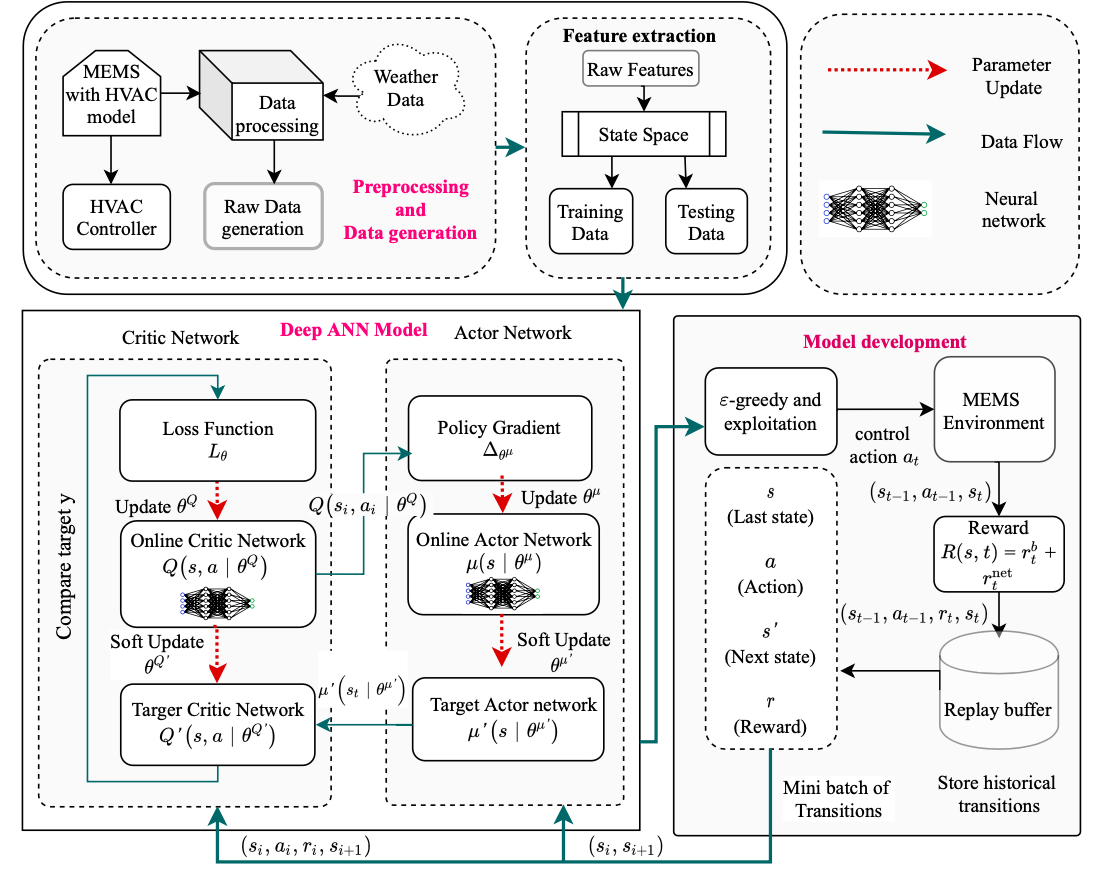} 
% }
\caption{DDPG Based HVAC-Aware Microgrid Control Framework.} 
%\vspace*{-0.4cm}
\label{cost}
\end{figure*}

\section{DDPG Based HVAC-Aware Microgrid Control Algorithm}
As discussed earlier, for solving this MDP problem, we propose a DDPG-based algorithm instead of DQN \cite{DQN}.
% At every step, the simulator feeds the state data to the controller and then the controller returns the battery charge/discharge power and air mass flow of every room by executing the DDPG model. The simulator adds up the battery charge/discharge power of every room to find the net charge/discharge power. Using the new air mass flow rates, a simulation step is then completed. The newly obtained temperatures of the rooms are sent to the DRL framework, which computes the reward and trains the model. These new temperatures are also used for the next step of the DDPG-DRL based controller to obtain the charge/discharge power of the batter and air mass flow rates of every room for the very next step. This cycle continues for a fixed number of episodes or until convergence.
% \vspace{0.6 mm}
\quad In the proposed Deep neural network model, the agent learns how the changes in MPC-based HVAC controller parameters affect the total power consumption of the building.
% Due to available noises and disturbances, it is difficult to obtain these transition probabilities accurately. For this reason Q-learning methods could be used. For this case with continuous system states, a function approximator could be adopted to evaluate Q-function where a neural network with weight $\theta$ is adopted as the non-linear function approximator which can be refered as Q-network. In \cite{DQN}, a deep Q-network (DQN) algorithm was proposed. However, DQN cannot be directly applied to the problem with continuous action spaces since it discretises the action space and lead to an explosion of the number of actions \cite{bd1}. As a result, decreased performance, decreased computational efficiency, and requirement of more training data would be observed \cite{EE}\cite{dim} \cite{policy}. We propose a Deep Deterministic Policy Gradients (DDPG) based algorithm for solving this issue.
% \vspace{0.5 mm}
\begin{algorithm}\footnotesize
\DontPrintSemicolon

  Initialize online critic network $C:Q(s_t, a_t|\theta^Q)$ and actor network $A:\mu(s_t | \theta^\mu)$ with random weights $\theta^Q$ and $\theta^\mu$ respectively.
  
  Initialize the target critic network $C':Q'(s_t,a_t |\theta^{Q'})$ and target actor network $A':\mu'(s_t|\theta^ {\mu'})$ with $\theta^{Q'} \leftarrow \theta^Q$ and  $\theta^{\mu'} \leftarrow \theta^\mu$ respectively.
  
  Initialise replay buffer $B$ of size $N$.
  
  \For{episode m=1,2,.......,M}
  { Obtain the initial system states $s_0$ for a random day in the training set.
 
  Initialize a random exploration noise $n_t$.
  
  \For{time step(in hours) t=1,2,......,T}
  {Obtain control action $a_t$ according to equation\eqref{noise} 
  
  Execute this control action $a_t$ in the environment, observe the reward $r_t$ after transition to the new system state $s_{t+1}$ at the end of time slot $t$.
  
  Store the experience $(s_t,a_t,r_t,s_{t+1})$ in the replay buffer $B$.
  
  Randomly sample a mini-batch of $K$ transitions $(s_i,a_i,r_i,s_{i+1})$ from replay buffer $B$, $ 1 \leq i \leq K $.
  
  Estimate reward for each sampled transition using equation \eqref{rewardmod}
  
  Update $C$ using equation \eqref{loss} over the sampled mini-batch.
  
  Update $A$ using sampled policy gradient as shown in equation \eqref{gradient}
  
  Update target networks $C'$ and $A'$ using equations \eqref{critic} and \eqref{actor} respectively.
  
  \textbf{end}}
  
\textbf{end}}

\caption{Training Procedure with DDPG}
\end{algorithm}
The detailed DDPG based training procedure is described in Algorithm 1. The DDPG based policy adopts an actor-critic framework based on Deterministic Policy Gradient (DPG) \cite{DPG}. The actor accepts the current state and predicts the actions as per our action space. It is trained using a mean squared error (MSE) loss. The output of the critic network is scalar, and it predicts the reward. It tries to map the reward to the reward function that we have provided for the environment. This network too is trained using MSE loss. The actor-network is defined as $ a_t=\mu(s_t | \theta^\mu)$, where $a_t$ represents the control action, $s_t$ is the system state and $\theta^\mu$ represents the weights of the actor-network. The critic network is defined as $Q(s_t, a_t|\theta^Q)$, where $a_t$ is the specified control action by the actor-network and $\theta^Q$ represents the weights of the critic network. The action-value function, $Q(s_t, a_t|\theta^Q)$ defines the expected reward by taking action $a_t$ at state $s_t$ by following the policy. 

DDPG uses a replay buffer to perform a random sampling of experiences. It updates the network parameters accordingly. All the features of the experiences are saved $(s_i,a_i,r_i,s_{i+1})$ and stored in a finite-sized (say $N$) cache referred as a “replay buffer.” Then, random mini-batches of experiences are sampled while the value and policy networks are updated.
The training for the DDPG network while interacting with the HVAC-aware MEMS environment. At each time slot t, we first obtain the system states, and send these states as input to the policy network, which finally give the output as the control action. We need to explore the state space so that the policy does not converge to local optimal solutions during the training. Therefore, for exploration we add a random noise to the obtained control action,
\begin{equation}\label{noise}
 a_t=\mu(s_t | \theta^\mu)+ n_t  
\end{equation}
where $n_t$ is the random noise. In our work, we use an Ornstein-Uhlenbeck process \cite{noise} for generating the noise, $n_t$, for exploration. After that the obtained $a_t$ is applied to our system. At the end of time slot $t$, the new thermal state $s_{t+1}$ is obtained and then the total reward $r_t$ during that time slot. Each transition $(s_t,a_t,r_t,s_{t+1})$ will be stored in the replay buffer B for training the actor network and the policy network. After that, we will randomly sample N transitions from the replay buffer for training the
network. For each transition $(s_i,a_i,r_i,s_{i+1}) \in N$, the estimated reward is calculated as follows where $\Gamma$ is the discount factor for model update,
\begin{equation} \label{rewardmod}
    r_i = r_i + \Gamma Q'(s_{i + 1}, \mu '(s_{i + 1} | \theta^{\mu '}) | \theta^{Q'})
\end{equation}
Then the critic network is updated by minimising the MSE error(i.e. loss) between the estimated reward ($r_0$ calculated from Eq.\eqref{rewardmod})
and the reward predicted by the critic network $Q(s_i, a_i)$ over the sampled mini-batch as shown below,
\begin{equation} \label{loss}
    Loss (L_{\theta}) = (1/N) \sum_i (y_i - Q(s_i, a_i|\theta^Q))^2
\end{equation}
\vspace{0.6mm}
After that, the actor network is updated using the policy gradient \cite{policy} as follows,
\begin{equation} \label{gradient}
    \Delta_{\theta^\mu} = \Delta_{a} Q(s,a) \Delta_{\theta^\mu}\mu(s | \theta^\mu)
\end{equation}
Finally, the target critic and actor networks are updated using the following equations respectively,
\begin{equation}\label{critic}
    \theta^{Q'} = \tau\theta^Q + (1 - \tau) \theta^{Q'}    
\end{equation}
\begin{equation}\label{actor}
    \theta^{\mu'} = \tau\theta^\mu + (1 - \tau) \theta^{\mu'}    
\end{equation}
After the completion of the training, only the actor network can make control action.
The parameters i.e. the weights of the actor and critic networks become fixed after the training as illustrated in Algorithm 1. Now the trained DDPG model can be deployed to provide the real-time energy management strategy for the HVAC-aware microgrid system. In Algorithm 2, the decision-making process is outlined. For a certain day, the MEMS agent observes the current system state $s_t$, and sets the control actions according to the learned policy by DDPG at each time step t.  For the partially known system model, the actions are battery charging/discharging power and tuning the parameters of the system. The actions are battery charging/discharging power and air mass flow rate for the completely unknown system model. Then the obtained control actions are mapped to energy schedules of the physical different units in the MEMS (Section III) along with HVAC of the building.
\vspace{0.8 mm}
\begin{algorithm}[h]\footnotesize
\DontPrintSemicolon

  Load the parameter $\theta^\mu$ of the online actor-network after the training procedure as illustrated in Algorithm 1.
  
  \For{day =1,2,.......,$N^{test}$}
  { Obtain the initial system states $s_0$ for the day.
 
  \For{time step(in hours) t=1,2,......,T}
  {Obtain control action $a_t$ according to equation\eqref{noise}.
  
  Execute this control action $a_t$ in the MEMS environment, observe the reward $r_t$ after transition to the new system state $s_{t+1}$ at the end of time slot $t$. Use $s_{t+1}$ for finding $a_{t+1}$.
 
  \textbf{end}}

\textbf{end}}

\caption{DDPG Based HVAC-aware MEMS Control Strategy}
\end{algorithm}
\begin{figure*}[h]
\centering\vspace*{-0.1in}
\resizebox{\linewidth}{!}{
\includegraphics[scale=2]{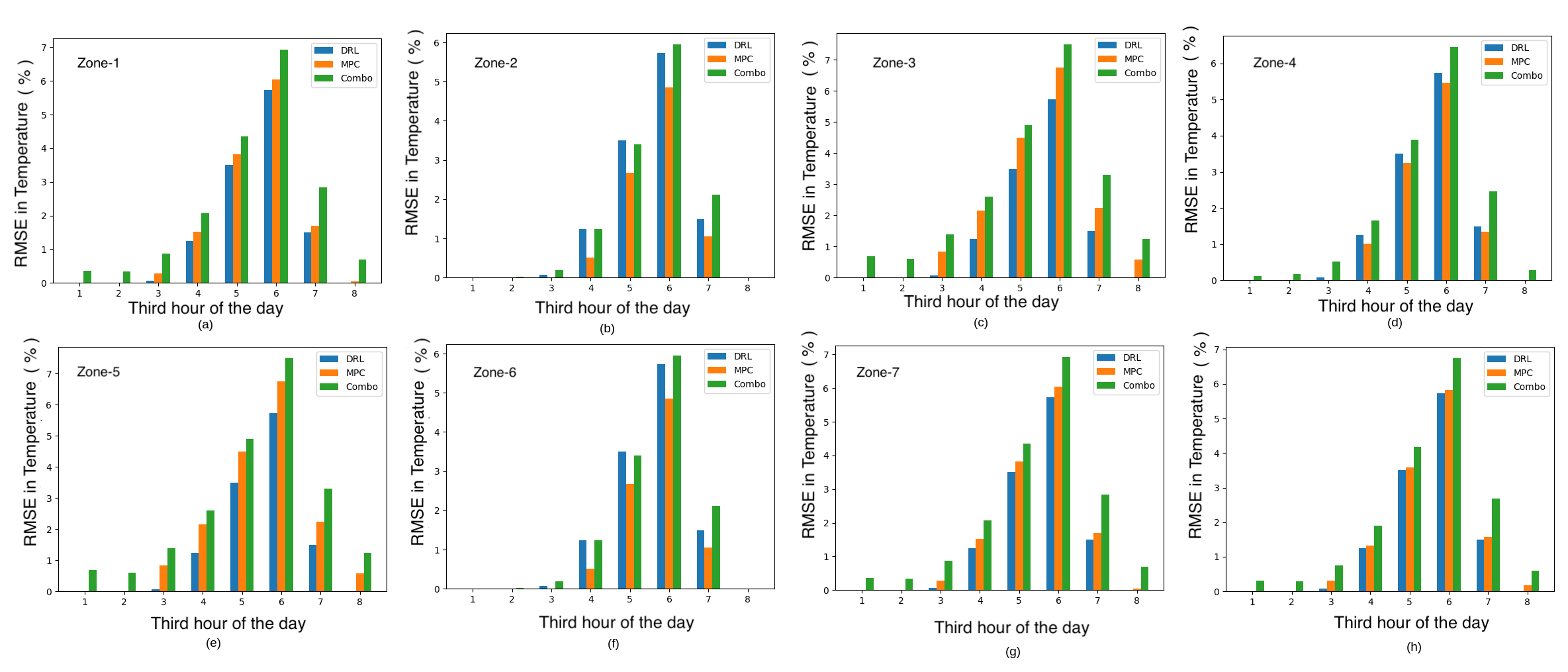}
}
\caption{RMSE (\%) in Temperature: subfigures a-g for each zone, subfigure h for all zones.}\label{rmse} \vspace*{-0.1in}
\end{figure*}
\section{Simulation Setup}
We implement the MEMS along with the building loads with the HVAC system for capturing the dynamics of the overall system. The main components of the system are illustrated in Figure-\ref{fig:framework}. We use OpenModelica \cite{openmodelica} platform to simulate the HVAC system and the thermal dynamics in the building. In the simulations, we have assumed that there is no change in building occupancy or its thermal load due to radiative forcing.     
% The training and evaluation of the DDPG model are done in a real-world MEMS environment developed. %As we know, power generation in microgrid and power consumption of the building loads largely depend on the weather condition. 
% We have used the weather data \cite{wdata1} of Kolkata, India for the experimental purpose. The electricity buy price follows the real-time price structure provided in \cite{realprice}.
The experiments are conducted on a Linux system, with  8-core $3.4$ GHz each and $8$ GiB of RAM. These experiments are implemented using Pytorch $1.5.1$ \cite{NIPS2019_9015}. 
We start with an MPC based controller for HVAC (designed using Matlab MPC toolbox). In this simulation, we worked with {\textit{seven}} zones. Zones represent different segments of a building that have negligible thermal coupling. This controller takes two variables as input - a state vector for the temperature of each zone, and the scalar parameter $w_c$ (as defined earlier),  which is set manually (to an initial value of $0.5$). The output produced by the controller is the air mass flow rate for each zone. 
% Here, the objective of eMPC is to minimise energy consumption and maintain the comfort factor within the desired range. This implies the minimisation of the difference between present zone temperature and it's desired temperature.  
% At each control instant, the predictive controller solves the optimisation problem as given in Equation 11 of \cite{d5}. We design a microgrid (as described in the section-IIA), connected to the utility grid via PCC, with the buildings as loads. The training and evaluation of the DDPG model are done in a real-world microgrid energy management environment developed. %As power generation in microgrid and power consumption of the building loads largely depend on the weather, 
We have used the weather data \cite{wdata1} of Kolkata, India, for the experiment. The electricity buy price for our system follows the real-time price structure provided in \cite{realprice}.
  
For the DDPG implementation, we use two DNNs (i.e. online and target) for the actor and the critic, respectively. The actor network and the critic network have two hidden layers with 128 neurons in each layer. We use the ReLU activation function and batch normalisation in each layer. We adopt Adam \cite{adam} for gradient-based optimization and the learning rate is 0.001. We use the target network’s soft update parameter, $\tau$ as 0.001 and the batch size 128. The duration of each time slot is 30 minutes, and each episode consists of 48 time slots. The initial exploration noise scale $\epsilon$ is taken as 0.35. Model parameter $\gamma$ is set to 0.05, $\sigma$ to 0.3, and dmax is set to 0.9, along with cmax. We take the value of $\kappa_{up}$ as well as $\kappa_{low}$ as 0.5. Here, $\lambda$ is assigned a value of 0.15. The partially known model(Combo) is trained for 50 epochs, whereas the completely unknown model(DRL) is trained for 100 epochs.

\section{Results \& Discussions}

% \begin{figure}
% \centering\vspace*{-0.1in}
% \resizebox{\linewidth}{!}{
% \includegraphics[scale=2]{Result/power_avg.png}
% }
% \caption{Average Power consumed (in KW) across all zones.}\label{power}
% \end{figure}
% \SayanEnd
% In Figure-\ref{temp}. we have the temperature profiles in all zones. Along with it, Figure-\ref{rmse} presents the Root Means Square Error (RMSE) of temperature with respect to the desired temperature for all the zones.  They are computed at intervals of 3 hours, leading to eight such points in 24 hours.  As expected, the trend here is similar to that of power usage.
% In figure-\ref{power}, we have the average power consumption flow of all the zones of the building. We have taken a random test day to show the power and temperature profiles of the building. The average power used helps us in getting a clear picture of the scenario and also draws a comparison between the three models. %It can be seen that the performance of the DRL and MPC models are comparable and similar. However, it is to be noted that the partially known model does not give a better result than the other two.  %Just as in the simulation steps, the day has been divided into 48 parts, at half an hour intervals.
%\SayanStart
\begin{figure*}
\centering\vspace*{-0.1in}
\resizebox{\linewidth}{!}{
\includegraphics[scale=2]{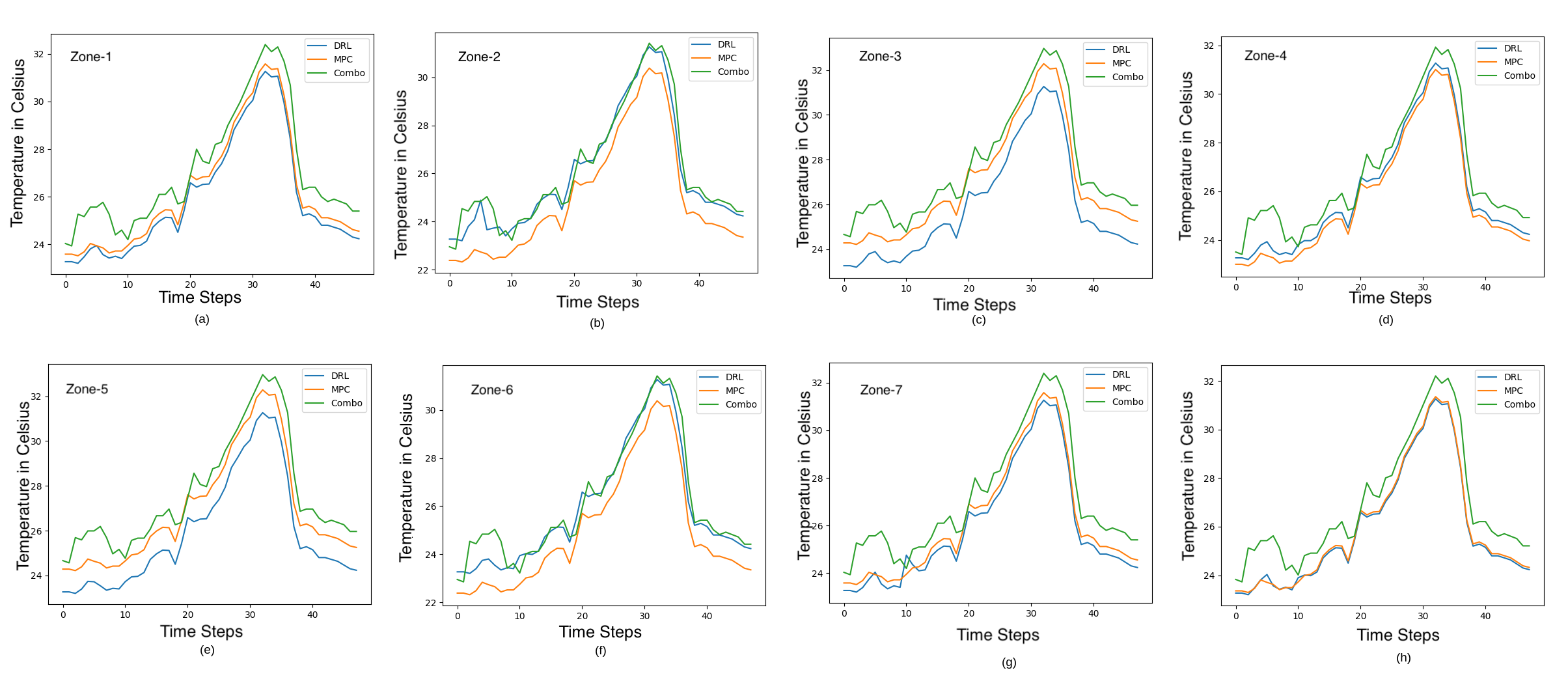}
}
\caption{Subfigures (a-g) denote the Temperature for each zone, subfigure (h) denotes the Average Temperature for all zones.}\label{temp} %\vspace*{-0.1in}
\end{figure*}\vspace{0.5mm}

\vspace{1mm}
% \begin{figure}[h]
% \centering\vspace*{-0.1in}
% \resizebox{\linewidth}{!}{
% \includegraphics[scale=2]{Result/power_avg.png}
% }
% \caption{Average Power consumed (in KW) in all zones.}\label{power}
% \end{figure}\vspace{1mm}

We present comparative plots among the three modes of abstraction. Figure-\ref{temp} gives the average temperature for each zone, along with an overall comparison. Figure-\ref{rmse} shows the Root Mean Square Error of temperature with respect to the desired temperatures for each zone. 
%  Figure \ref{power} outlines the power usage using the different control policies.
The average power usage of all zones has a similar curve as Figure-\ref{temp}.
Here the DRL model ({\em scenario-3}) consistently performs at par (and even better at times) with that of the MPC model ({\em scenario-1}). However, the MPC model optimised with a DRL model (the combo scheme, i.e. {\em scenario-2}) does not give better results in comparison with the other two. As mentioned earlier, the day has been divided into 48 parts at half an hour intervals. All three plots validate similar inference. RMSE is computed at intervals of 3 hours, leading to eight such points in a 24 hour period.

% We note that the zone temperatures obtained in the DRL model feature less standard deviation, which is a positive result owing to the fact that drastic changes in temperature, even within the comfortable range of temperature, is undesirable. This is because the DRL  learns from experience and avoids unnecessary fluctuations.  %Moreover, the DRL model being a complete black box provides us with the ease of modifying the environment without making any significant change to the model itself. 
%The models used in this work were all trained on local CPUs, and thus it did not have the overhead of GPU computation resources. 

Table-\ref{tab:results} provides the quantitative results of our work at various levels of abstraction of the system. The word ``Combo" is used to designate the partially known model. As evident from the values, the mean values of the completely unknown (DRL) model are comparable (and at times better) w.r.t. those of the completely known (MPC) model.
Hence, in this specific case, even in the presence of abstractions, AI-based learning algorithms can successfully mimic the scenario. The Table-\ref{tab:results} also highlights that the standard deviation of average temperature is less in the case of DRL compared to MPC. This is a positive result owing to the fact that drastic changes in temperature, even within the comfortable range of temperature, is undesirable. This is because the DRL  learns from experience and avoids unnecessary fluctuations. However, the standard deviation is less for the partially known model though the model's power consumption was significantly higher. 
% \sd{figures not clear}
% \vspace{1mm}
% \begin{figure}[H]
% \centering\vspace*{-0.1in}
% \resizebox{\linewidth}{!}{
% \includegraphics[scale=2]{figs/reward_all.png}
% }
% \caption{Subfigures a-g: Reward for DRL model for each zone; Subfigure h: Reward for DRL-based controller model for all zones }\label{reward} %\vspace*{-0.1in}
% \end{figure}\vspace{1mm}
% In the completely known model (mentioned as MPC in the figures), the minimum exploration probability  $\epsilon$ is 0.1 and the system parameters like solar radiation power, non-shiftable power demand, ambient temperature, and electricity price, are varying in each episode. As a result, the episode reward fluctuates within a small range. To show the changing curve of rewards more clearly, we provide the average value of the past 5 episodes. In figure-\ref{reward}, it can be found that the average reward generally increases and becomes more and more stable.

\vspace{1mm}

\begin{table}[!htp]\centering
\small
\begin{tabular}{lcccc}\toprule
\multirow{2}{*}{} &\multirow{2}{*}{\textbf{Power Consumption }} &\multicolumn{2}{c}{\textbf{Temperature }($^{\circ}$C)} \\\cmidrule{3-4}
&(kW)&\textbf{Average} &\textbf{RMSE} \\\midrule
&\multicolumn{3}{c}{\textbf{Mean}} \\ \midrule
MPC &  1.496 &  25.971 &  1.602 \\
DRL &  1.384 &  25.907 &  1.506 \\
Combo &  2.035 &  26.885 &  2.180 \\ \midrule
% \scriptsize
&\multicolumn{3}{c}{\textbf{Standard Deviation}} \\ \midrule
% \normalsize
MPC &4.170 &6.025 &3.840 \\
DRL &4.208 &5.909 &3.853 \\
Combo &5.199 &5.870 &4.591 \\
\bottomrule
\end{tabular}
\caption{Mean and Standard Deviation of Power Consumption and Temperature for various levels of abstractions of the system model}\label{tab:results}
\end{table}
Adapting an MPC model for a new building environment requires additional parameters while incorporating precise environment features into the model, leading to an increase in the chances of errors. However, using DRL, an existing model can be retrained and installed, which leads to saving efforts and cost towards deployment. 
\section{Conclusion}
In this work, we attempt to develop two levels of abstraction for the completely known model of MEMS. The first level of abstraction deals with an MPC-based method amalgamated with DRL. The DRL tunes certain parameters of the MPC model instead of it being directly derived from the environment. The next layer of abstraction replaces the completely known model with a black-box model, which is completely simulated using a DRL architecture. We find that this model performs at par with our specific test cases and exhibits less standard deviation w.r.t. the MPC-based MEMS model. Using a DRL model has certain design advantages leading to efficient deployment. One of the future work would be using Twin Delayed Deep Deterministic Policy Gradients (TD3)\cite{TD3} instead of DDPG in our problem.

\bibliographystyle{plain}
\bibliography{bib}

\end{document}